\documentclass[11pt]{article}
\usepackage{hyperref}
\pdfoutput=1

\begin{document}
\title{Large--eddy simulations of Richtmyer--Meshkov instability in a converging geometry}
\author{Manuel Lombardini$^1$ \& Ralf Deiterding$^2$ \\
\\\vspace{6pt} $^1$ Graduate Aeronautical Laboratories, \\ California Institute of Technology, Pasadena, CA 91125, USA \\
\\\vspace{6pt} $^2$ Computational Science and Mathematics Division, \\ Oak Ridge National Laboratory, Oak Ridge, TN 37831-6367, USA}
\maketitle

%% The abstract (in this file, and that submitted as text to arXiv) should include the exact phrase
%% "fluid dynamics video" or "fluid dynamics videos"
\begin{abstract}
The Richtmyer-Meshkov instability (RMI) refers to the baroclinic generation of vorticity at a perturbed density interface when impacted by a shock wave. It is often thought of as the impulsive limit of the Rayleigh-Taylor instability (RTI). The fluid dynamics video \href{http://hdl.handle.net/1813/14106}{``large-eddy simulations (LES) of RMI in a converging geometry''} shows the mixing of materials resulting from the interaction of an imploding cylindrical shock wave with a concentric perturbed interface that separates outside light gas from heavy gas (initially 5 times denser) inside a wedge. At the initial impact, the incident shock Mach number is either 1.3 or 2.0. The present canonical simulations support recent interests on compressible turbulent mixing in converging geometries relevant to both inertial confinement fusion and core-collapse supernovae dynamics.
\end{abstract}

\section{Flow configuration}

A converging cylindrical shock initially impacts a perturbed cylindrically-shaped density interface that separates (light) air from (heavy) SF$_6$ (density ratio of 5), both at rest at a radial position $R_0=1$m. Pressure and temperature are initially continuous across the interface, and set to 23kPa and 286K, similarly to the experiments of Vetter \& Sturtevant in plane geometry~\cite{vetter1995}. Two incident shock Mach numbers $M_I=1.3$ and 2.0, defined just before shock-interface refraction, are tested.

The flow is confined to a 90$^{\circ}$ wedge. We impose periodic boundary conditions in the $z$-direction of the cylinder axis (axial extent $L_z=1$m). Zero-gradient boundary conditions are employed over an outer cylindrical boundary. It is assumed that shock wave-boundary layer interaction does not play a dominant role in the growth of the mixing layer and slip boundary conditions can be applied at the horizontal and vertical reflecting walls. An inner cylindrical wall is also used to regularize the apex. 

Initial irregularities in the density interface form the misalignment between density and pressure gradients required to initiate RMI. A second RMI occurs after the initial shock has converged down the wedge, reflected off the axis and reshocks the distorted interface. Reshock interactions of decreasing intensity follow successively. Due to the converging geometry, the accelerated or decelerated motion of the interface also generates RTI. As the fluids start to interpenetrate, secondary Kelvin-Helmholtz instabilities develop along the sides of the fingering structures. 

\section{Numerical approach}

The reshock produces a large dynamical range of turbulent scales, requiring the techniques of LES. We employed the stretched-vortex subgrid-scale model of turbulent and scalar transport based on an explicit structural modeling of small-scale dynamics, as developed by Pullin and co-workers. The imploding nature of the flow is particularly suitable for the use of adaptive mesh refinement (AMR) provided by the parallel block-structured AMR framework AMROC. The Favre-filtered Navier-Stokes equations are solved on each Cartesian uniform subgrid of the mesh hierarchy. A weighted, essentially non-oscillatory scheme is used to capture discontinuities but reverts to a low-numerical dissipation, explicit, tuned center-difference stencil in the smooth or turbulent flow regions, optimal for the functioning of our explicit LES (See~\cite{pantano2007} for more details). 

The $(x,y,z)$-domain is discretized with $95\times95\times64$ cubic cells on the base grid with three additional levels of refinement based on the local density gradient, reducing the computational expenses compared to the equivalent finest unigrid $760 \times760\times512$ problem. The initial perturbation wavelength is 100 times larger than the finest grid size, which is itself about 100 times larger than the estimated smallest physical scale in the flow - the Kolmogorov scale - where the viscous dissipation of energy occurs.

\section{Video}

The video \href{http://hdl.handle.net/1813/14106}{``LES of RMI in a converging geometry''} shows four animations for each incident shock Mach number:
\begin{enumerate}

\item magnitude of Schlieren density contours in perspective view;

\item iso-surfaces of heavy-fluid mass fraction $Y_{\textup{SF}_6}$ in perspective view (between iso-surfaces $Y_{\textup{SF}_6}=0.1$ and 0.9);

\item levels of Schlieren density and heavy-fluid mass fraction in azimuthal and axial plane slices through the mixing region; 

\item AMR levels of refinement in perspective view. 

\end{enumerate}
For the case $M_I=1.3$ (resp. 2.0), each animation shows the flow evolution for about 25 $\times 10^{-3}$s (resp. 15 $\times 10^{-3}$s), and is played at a speed about 800 (resp. 1,100) times slower than in experimental conditions. Data are sampled about every 0.5$\times 10^{-3}$s.

The first animation successively shows the location of the incident shock before impacting the initial perturbed interface; growing mushroom-like structures characteristic of baroclinic instabilities as the interface accelerates towards the axis of the cylinder, following the transmitted imploding shock; the reflection of the imploding wave off the axis and the first reshock interaction with the interface; the subsequent interpenetration of fluids and the development of a wide range of scales. The second movie portrays the mixing of light fluid (green) with heavy fluid (red) by displaying isosurfaces of heavy-fluid mass fraction. Both Schlieren density and heavy-fluid mass fraction representations are displayed together in the third sequence, where azimuthal and axial plane slices through the mixing region are set side by side, showing the contrast between azimuthal and axial modes. The fourth animation indicates the dynamics of each level of grid-refinement and suggests the computational savings achieved thanks to the AMR strategy. Strong incident shock strengths (\textit{e.g.} $M_I=2.0$) focus and compress the flow further down towards the axis, reshock interactions occur closer to the axis with greater intensity, leading to a stronger turbulent mixing.

\section{Acknowledgment}

This work was supported by the ASC program of the Department of Energy under subcontract no. B341492 of DoE contract W-7405-ENG-48. The simulations were performed at the Center for Advanced Computing Research (CACR) at the California Institute of Technology.


\begin{thebibliography}{1}

\bibitem{vetter1995}
M.~Vetter and B.~Sturtevant.
\newblock Experiments on the {R}ichtmyer--{M}eshkov instability of an
  air/{SF}$_6$ interface.
\newblock {\em Shock Waves}, 4:247--252, 1995.

\bibitem{pantano2007}
C.~Pantano, R.~Deiterding, D.~J. Hill, and D.~I. Pullin.
\newblock A low numerical dissipation patch-based adaptive mesh refinement
  method for large-eddy simulation of compressible flows.
\newblock {\em J. Comput. Phys.}, 221(1):63--87, 2007.

\end{thebibliography}
\end{document}